# Mechanical-scan-free and multi-color super-resolution imaging with diffractive spot array illumination

Ning Xu[1], Sarah E. Bohndiek[2,3], Zexing Li[4], Cilong Zhang[1], and Qiaofeng Tan[1]*

[1] *State Key Laboratory of Precision Measurement Technology and Instruments, Department of Precision Instrument, Tsinghua University, Beijing 100084, China*
[2] *Department of Physics, Cavendish Laboratory, University of Cambridge, JJ Thomson Avenue, Cambridge, CB3 0HE, UK*
[3] *Cancer Research UK Cambridge Institute, University of Cambridge, Robinson Way, Cambridge, CB2 0RE, UK*
[4] *Department of Pure Mathematics and Mathematical Statistics, University of Cambridge, Wilberforce Road, Cambridge, CB3 0WB, UK*

*Corresponding author: tanqf@mail.tsinghua.edu.cn

## Abstract

Point-scanning microscopy approaches are transforming super-resolution imaging. Despite achieving parallel high-speed imaging using multifocal techniques, efficient multi-color capability with high-quality illumination is currently lacking. In this paper, we present for the first time Mechanical-scan-free and multi-Color Super-resolution Microscopy (MCoSM) by spot array illumination, which enables mechanical-scan-free super-resolution imaging with adjustable resolution and field of view (FoV) based on spatial light modulators (SLMs). Through 100s-10,000s super-resolution spot illumination with different FoV for imaging, we demonstrate the adjustable capacity of MCoSM. MCoSM extends current spectral imaging capabilities through a time-sharing process of different color illumination with phase-shifting scanning, while retaining the spatial flexibility of super-resolution imaging with diffractive spot array illumination. To showcase the prospects for further combining MCoSM with multi-color imaging, we also perform spectral unmixing (four-colors) on images of fluorescent beads at high resolution. MCoSM provides a versatile platform for studying molecular interactions in complex samples at the nanoscale level.

## INTRODUCTION

Building on the success of confocal microscopy (CM) [1, 2], super-resolution microscopy is indispensable for investigating nanoscale biological structures and interactions in biomedical research. [3, 4]. Using multiple distinct fluorescent labels to realize multi-color super-resolution imaging enables researchers to study spatial relationships between different molecular processes within cells [5]. Traditional point-scanning super-resolution microscopy can be challenging when applied for multi-color imaging due to the long data-acquisition time, which normally exacerbates photobleaching. In addition, movement of the microscope stage typically requires mechanical adjustments, which during refocusing can perturb the sample although is sometimes avoided by remote focusing [6, 7]. These obstacles have led to difficulty in performing multi-color fluorescent imaging with nanometer-scale precision in biological samples.

To reduce the data-acquisition time, multifocality has been explored in super-resolution microscopy to extend the total composite field of view (FoV) [8]. If the conventional single spot is replaced with a large enough spot array, the data-acquisition time can be significantly reduced with only a few steps of sampling. There is a direct relationship between the FoV and the spot size [9], given by

$$\text{FoV} \sim \frac{\text{FN}}{M_{\text{sys}}} = \alpha \cdot \frac{1.22\lambda}{\text{NA}} \frac{N}{M_{\text{sys}}} \quad , \tag{1}$$

where FN is the field number, NA is the numerical aperture of the objective, $\lambda$ is the wavelength of the incident beam, $0 < \alpha < 1$ is the ratio of spot resolution to Airy spot, $N$ and $M_{\text{sys}}$ are respectively the number of spots in the array and the magnification of the system. For a 50×50 spot array with a resolution of 70% of the Airy spot (0.7Airy), each position collects an area of about 1750 times increased with a 30% improvement in lateral resolution compared with single Airy spot illumination. Therefore, the spot array can reduce scan times by 2-3 orders of magnitude compared to a single spot illumination, while improving the resolution.

Nonetheless, creating super-resolution spot arrays beyond the diffraction limit remains challenging. Initially proposed by Toraldo di Francia [10], the process of focusing a single super-resolution spot beyond the Abbe-Rayleigh diffraction limit can be realized by wavefront shaping [11-14] or computational deconvolution [15, 16], or via some combination of these [17, 18]. Flexible phase distributions can generate a range of modulated intensity distributions by harnessing the power and affordability of diffractive optics for altogether new illumination strategies within optical microscopy and nanoscopy, making mass production feasible assisted by nano-printing, with the potential cost per element possibly reducing to just a few dollars. Super-resolution spot arrays can achieve high resolution by diffractive optics but at the expense of extremely low light efficiency due to high intensity sidelobes and speckles, which limits the applications in nanoscale imaging [19-21]. In our previous work, we experimentally generated 3×3 super-resolution spot arrays by diffractive optical elements [21], however, realizing higher FoV by increasing the number of spots with higher resolution in state-of-the-art microscopy is highly demanding in practice.

To achieve super-resolution multi-color acquisitions, mechanical perturbations and the uniformity of the spot array are the main challenges in achieving adjustable imaging. Typically, multi-color information is obtained using multi-wavelength illumination tuning in combination with a time-sharing scheme to obtain organelle information. Notably, wavelength-selective phase shift has been adopted in digital holography [22] and three-dimensional imaging [23, 24] to replace mechanically moving components. Various organelles can be imaged by phase-shifting multiplexing with high efficiency and avoiding the need to mechanically perturb the sample.

In this paper, we propose a method to design and build Mechanical-scan-free and multi-Color Super-resolution Microscopy (MCoSM) for mechanical-scan-free imaging with adjustable diffractive spot arrays using spatial light modulators (SLMs). As a proof-of-concept study, we performed imaging of fluorescent beads and biological tissues using MCoSM, where the maximum scanning range and minimum step size are 210.4 μm and 79.1 nm in $x$ and $y$ axes, respectively. MCoSM experimentally achieves four-colors imaging, where the resolution is better than 96 nm using a 10×10 spot array with a resolution of 52% of the Airy spot (0.52Airy), advancing on prior art [25] and achieving simultaneous multi-color phase-shifting scanning via superposing and rapidly modulating the designed phase distribution on the SLM. The current solution has a great advantage in controlling the beam modulation and scanning process with the same SLM enabling high precision and rapid imaging is achieved for the respective wavelengths. As every phase distribution corresponds to a specific wavelength and position in space, the phase shift by the SLM can move the spot array to any location within its work region, thereby realizing mechanical-scan-free imaging.

## RESULTS

**Principles of MCoSM**

In our MCoSM implementation, the illumination is provided by super-resolution spot arrays, with the adjustable numbers of spots and spot sizes generated by SLMs. Multi-color imaging is achieved by automatically alternating simultaneous multimodal acquisitions (four independent channels) and phase-shifting lateral mosaicking. Successive excitation in different visible bands is provided by lasers at separate wavelengths (Fig. 1a). With the number of labels increasing, the incident wavelengths can be correspondingly increased. Every single wavelength is split into two by a beam splitter (BS) and the modulated transmitted beam is reflected off by SLM. After being reflected at the BS, the beam illuminates at the pupil plane of the objective, and subsequently the spot array is generated in the sample plane located at the focal plane of the objective to illuminate a target object. Tuning the power of the pulses and their time delay enables control of the excitation windows in a largely independent manner. This strategy provides efficient, simultaneous excitation of several fluorescent labels with phase-shifting. In other words, the spatial resolution and spectral channels of MCoSM can be easily altered by switching the parameters of the illumination spot array to meet the requirements (see Methods).

Super-resolution information is not directly obtained by the imaging system so must be reconstructed. The intensity of MCoSM on the detector can be analytically expressed as [9]

$$I(\xi,\eta)=\iint p_{\text{illu}}(u,v)t(u,v)p_{\text{imag}}(\xi-u,\eta-v)dudv, \qquad (2)$$

where $p_{\text{illu}}(u,v)$ and $p_{\text{imag}}(\xi,\eta)$ are respectively the PSFs of the illumination system and imaging system, and $t(u,v)$ is the intensity transmittance function of the sample. Previously, we have shown that super-resolution information can be obtained from the convolution of the imaging system in numerical simulation with a 3×3 or 5×5 spot array incidence when the NA of the illumination system equals that of the imaging objective, while the distance between the adjacent spot centers is twice the spot size [25]. To clarify the super-resolution information also can be obtained when the number of spots extended to $N^2$ ($N$>5), we firstly mathematical demonstrated the relative position and intensity of the spots were not affected by the convolution of the imaging system when the NA of the illumination system was equal to the imaging system (see Supplement note 1).

Realizing phase-shifting scanning depends on the tilted phase loaded on the SLM, which is used to replace the transverse mechanical-scan process. To scan the spot array along the $u$-plane or $v$-plane in the focal plane of the objective, one can vary the value of spatial frequency of the tilted phase term that controls the separation distance (or angle) between the -1st and 0th diffraction order, to determine the path of the phase-shifting (Fig. 1b). The $u$-plane and $v$-plane in the focal plane are independent of the phase-shifting control and the sizes of spot arrays should be smaller than the maximum scan range [24, 26, 27]. Small or high values of spatial frequency should be avoided, as they may cause the diffracted beams to overlap in space or resulting in no modulation of SLM [31]. Therefore, the minimum step size and maximum scan range are of great importance to decide the FoV and resolution of MCoSM (Fig. 1c). The minimum step size of phase shift $\delta_{u,\min}$ can be expressed as

$$\delta_{u,\min} = \frac{\lambda f_{L1}}{M_{sys}} f_{u,\min} = \frac{\lambda f_{L1}}{M_{sys} L_{u,\max} \Delta p} \ . \tag{3}$$

The minimum step size can achieve $\delta_{u,\min} = 79.1$ nm by setting $L_{u,\max} = 1920$ and $\lambda_1 = 405$ nm. The minimum step size can be further reduced by introducing an SLM with more pixels, as indicated in Eq. (3).

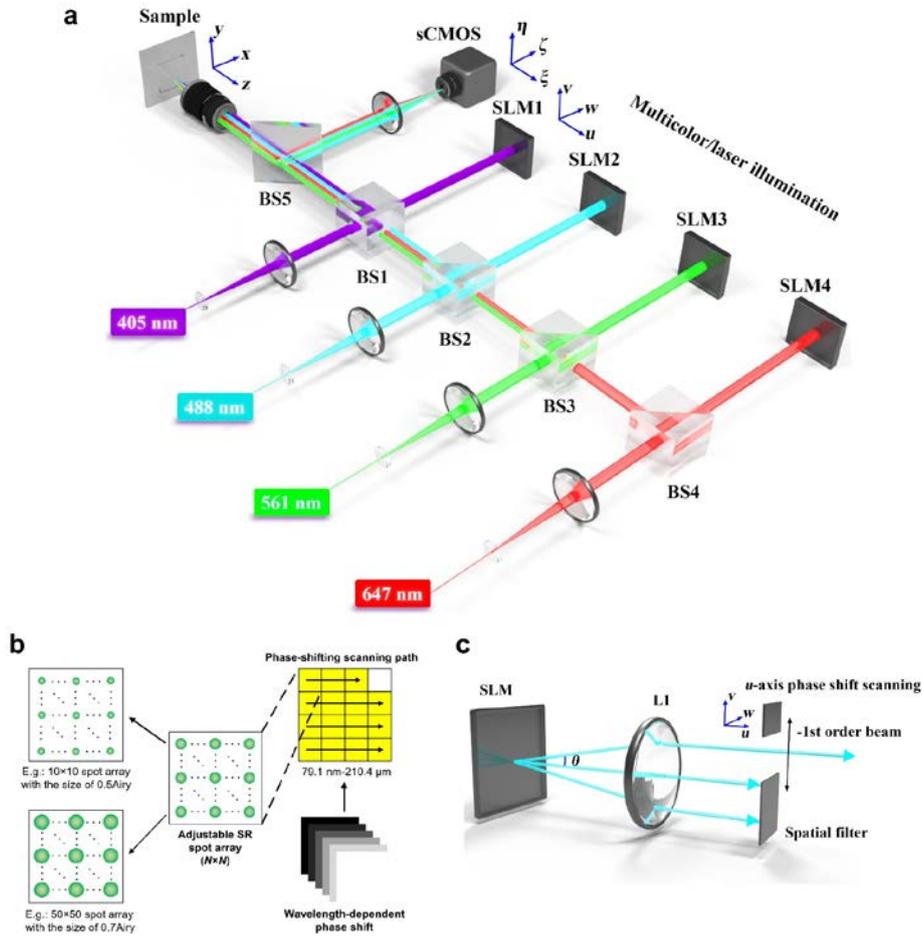

**Fig. 1. Optical design of MCoSM system and illumination phase-shifting scanning path. a** The system is assembled using 4 SLMs with plane wave incidence, to generate *N*×*N* super-resolution spot(s) on the sample, and an imaging detector, to relay super-resolution information. **b** Phase-shifting scanning path of the spot array, which is realized by adding different phase shift on SLMs. **c** Phase shift scanning in the focal plane of L1, where *u*-plane is realized via varying the spatial frequency $f_u$. $\theta$ is the diffraction angle between the 0th and -1st-order diffraction. BS: beam splitter, SLM: spatial light modulator, sCMOS: scientific complementary metal oxide semiconductor.

The maximum spatial frequency variation can be expressed as $f_{u,\max} = 2/(L_{u,\min}\Delta p)$, and the scan range can be expressed as

$$\delta_{u,\max} = \frac{\lambda f_{L1}}{M_{sys}} f_{u,\max} = \frac{2\lambda f_{L1}}{M_{sys} L_{u,\min} \Delta p} \ . \tag{4}$$

The maximum scan range ($\delta_{u,\max}$) can be achieved by 210.4 μm by setting $L_{u,\min}=2$ and $\lambda_4 = 561$ nm. These results indicate that the MCoSM has suitable scanning resolution for performing continuous phase shift. Similarly, the expression of the minimum step size ($\delta_{v,\min}$) and maximum scan range ($\delta_{v,\max}$) at *v*-plane are the same with Eqs. (3) and (4). Notably, aberration will affect the spot array near the maximum scan range, especially for large NA, which is practically constrained by the scan range (see Supplement note 2).

**Fluorescent beads phase-shifting imaging: Four-colors with 100 nm resolution**

To present the further spatial and spectral characteristics of MCoSM, we performed MCoSM in four-colors using fluorescent beads with a diameter of 100 nm. The beads were labeled with blue, green, red, and cyan fluorescent dyes corresponding to different wavelengths of excitation.

In this experiment, the theoretical diffraction-limited resolution was determined by the NA of the objective lens (NA=1.25) and emission wavelength. Here, a 10×10 spot array with a resolution of 0.52Airy was used as the illumination to scan the fluorescent beads (see Visualization 1). Imaging outputs qualitatively indicated resolution enhancement compared to the widefield image of the beads (Fig. 2). According to statistical analysis of quantitatively derived data from 15 different single beads, the average size of a single bead in MCoSM image was reduced to 96±9 nm compared to 197±22 nm in the widefield at 405 nm. The 197 nm size of the beads in the widefield image met the 198 nm theoretical diffraction-limited resolution well, and the much narrower 96 nm size in the MCoSM image proved that super-resolution imaging can be further extended when the NA is improved. The line profiles indicated that four-colors MCoSM was capable of resolving adjacent fluorescent beads and super-resolution can be achieved (Figs. 2b-2e).

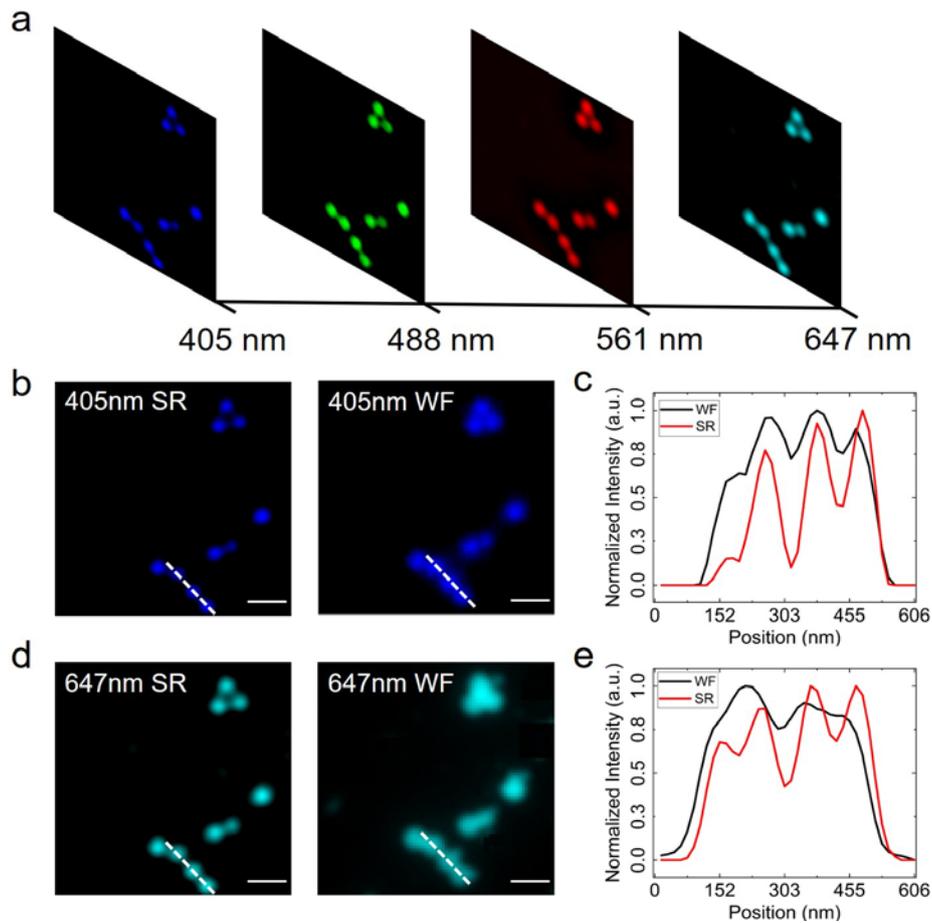

**Fig. 2. Imaging results of fluorescent beads. a** Four-colors fluorescent bead at the illumination wavelength of 405 nm, 488 nm, 561 nm, and 647 nm. **b** and **d** Pseudocolor MCoSM image and widefield image of 405 nm and 647 nm marked by the white dotted line. **c** and **e** Intensity profiles on the selected in **b** and **d**. Scale bar: 10 μm. (see Visualization 1)

**BPAE imaging: FoV up to 167.4 μm×167.4 μm with 0.52Airy resolution**

To demonstrate the spatial and spectral characteristics of the MCoSM in a biological sample, we imaged a three-colors stained slide from Bovine Pulmonary Artery Endothelial (BPAE) cells, with labels applied to the nucleus, cytoskeleton, and mitochondria using our home-built MCoSM prototype. Uploading six stacks phase distribution on the SLMs, we respectively reconstructed super-resolution images illuminated by 10×10 (*N*=10) and 50×50 (*N*=50) spot arrays with a resolution of 0.52Airy and 0.68Airy illumination, which contained 6 images in 2 FoVs with three-wavelength incidence ($\lambda_1$= 405nm, $\lambda_2$= 488nm, and $\lambda_3$= 561nm). The captured stacks (shown in Visualizations 2 and 3) were used as representative examples to illustrate the working principle of adjustable imaging, where the combined phase distributions were uploaded on the SLMs to scan the sample, reconstructing multi-color super-resolution images by spot arrays with phase-shifting scanning. The nucleus, cytoskeleton, and mitochondria are distinguished morphologically in the MCoSM images of two FoVs, presenting different fluorescence emission properties clearly in the spectral dimension with different trends (Figs 3a-3c; single wavelength reconstructed images with 10×10 spot arrays incidences in Fig. S7 of Supplement note 3). The distributions of nucleus, cytoskeleton, and mitochondria over the FoV could be visualized using reconstructed images at 461 nm (Fig. S7a), 512 nm (Fig. S7b), and 599 nm (Fig. S7c) created using a mosaic stitching algorithm. The signals of the three-colors were well separated with a negligible cross talk.

While extending the spectral dimension, the MCoSM technique also retains spatial super-resolution characteristics. In detail, we extracted the sample curves in the same FoV of the sample, showing that the MCoSM successfully resolved adjacent cytoskeleton, which would be mistaken as a single thick strand in widefield imaging results (Figs. 3a-3d). The Gaussian fitting result of the MCoSM intensity profile (Fig. 3d) showed that a lateral resolution of at least 193 nm was achieved using MCoSM at a wavelength of 488 nm while the corresponding theoretical diffraction-limited resolution was about 347 nm.

Nonetheless, the improved FoV of the proposed MCoSM technique does not come without a cost. MCoSM essentially transforms the majority of the data throughput, originally acquired from mechanical point-by-point manner, into phase-shifting scanning assisted with SLM. Considering the pixel size and stability of SLM, there is a tradeoff between the effective FoV and sampling interval. The FoV and sampling interval are determined by minimum step size (Eq. 3) and maximum scan range (Eq. 4). To achieve a high signal-to-noise ratio of the reconstructed image, there is some redundancy during data acquisition, especially for 10×10 super-resolution spot array illumination in the experiment. Hence, the effective FoVs of the multi-color super-resolution images are 61.4 µm×61.4 µm (Fig. 3a) and 167.4 µm×167.4 µm (Fig. 3b), respectively generated by 10×10 and 50×50 spot arrays with the size of 0.52Airy and 0.68Airy.

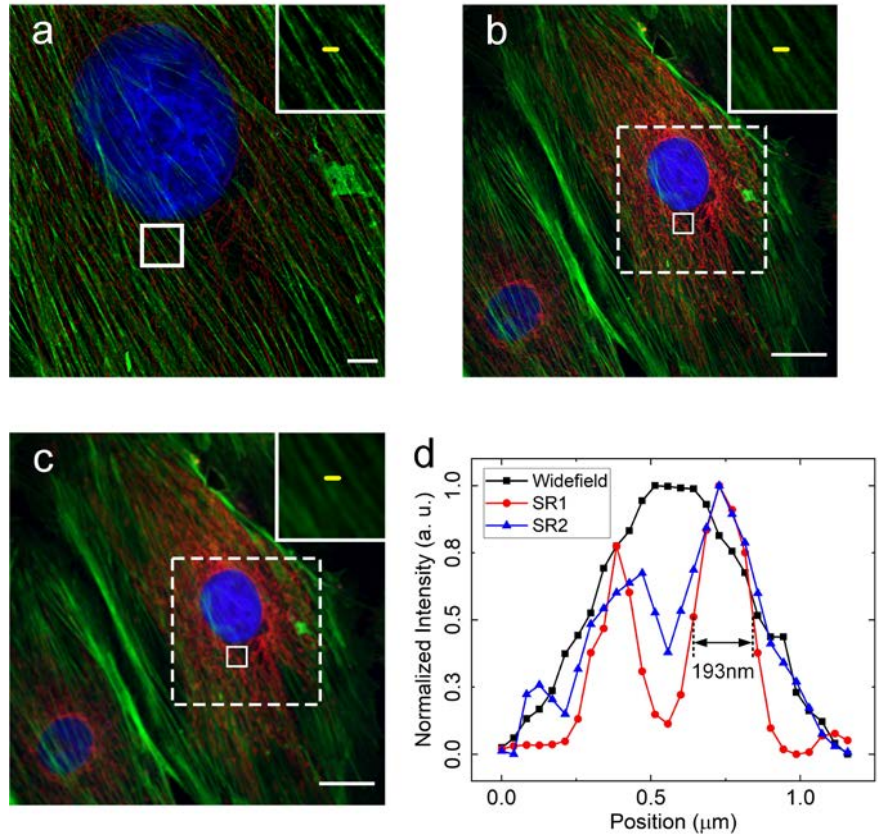

**Fig. 3. Spatial and spectral characteristics of the MCoSM imaging of BPAE.** The combination of MCoSM images at 461 nm, 512 nm, and 599 nm in the reconstructed images, describing the distribution of nucleus, cytoskeleton, and mitochondria over the FoV. **a** and **b** MCoSM images of organelles by 10×10 and 50×50 super-resolution spot array with resolution of 0.52Airy and 0.68Airy, respectively. **c** Widefield image of organelles with the same objective. (objective lens: 60×/0.90 NA). The region of interest (ROI) marked by the white solid rectangle was shown in the enlarged upper right boxed region. Comparisons of identical areas were marked with white dotted rectangle. **d** Intensity profiles of cytoskeleton on the selected line in **a** (SR1), **b** (SR2), and **c**. The white dotted box region in Figs. 3b and 3c have been selected to compare the FoV and spatial resolution with Fig. 3a. Scale bar: **a** 5 μm, **b** and **c** 25 μm. (see Visualizations 2 and 3).

## DISCUSSION AND CONCLUSION

MCoSM provides the capability of multi-color super-resolution microscopy with adjustable resolution and FoV without mechanical scanning for the first time. It achieves the spectral mechanical-scan-free imaging capability through a time-sharing process of different color illumination with phase-shifting, while retaining the spatial characteristic of super-resolution imaging with diffractive spot array illumination. We demonstrated the use of MCoSM for unmixing imaging from four-colors using fluorescent beads as a proof of concept retaining high resolution while opening new possibilities in practical applications through demonstration in a reference biological sample containing three-colors.

Although the concept of MCoSM has been verified here, there remains room for improvement in terms of scan range, speed, spatial characteristics and spectral sampling. Firstly, the maximum scan range and minimum step size are limited by the pixel size and pixel number of SLM. Secondly, as the phase-shifting is capable of arbitrary paths in space, all usual scanning strategies, e.g., raster, spiral, or Lissajous scanning trajectories, can be reasonably implemented on the SLM-based MCoSM system. The region within the spot array theoretically only requires the adjacent spot scanning to satisfy Nyquist sampling, hence subsequent improvements to the scanning method can be made to increase the imaging speed. In addition, the dwell time at each position of spot array on the scanning path can be controlled, realizing selective optical stimulation for biomedical samples. Boosting the speed of the phase shift from 10 Hz to 60 Hz, the maximum frame rate of SLM could also accelerate imaging and reduce fluorescence bleaching. Thirdly, spatial characteristics could be improved through further modification of the optimization algorithm i.e. achieving better spatial resolution, increasing the number and uniformity of spots, to obtain better performance in practice. It is already possible to generate spot arrays with higher spatial resolution and better uniformity if one can sacrifice the FoV [29]. Finally, MCoSM could also benefit from the development of metasurfaces [30], exploiting these for further miniaturizing the optical set-up.

In brief, our findings indicate that multi-color super-resolution microscopy can now overcome prior challenges in the field, enabling practical implementation in a variety of fluorescent, nanoscale biomedical imaging. Of note, it can be foreseen that an adjustable illumination strategy may provide a new research direction for performing complex light beam shaping for extending other nanoscopy methodologies such as multifocal structured illumination microscopy [31-33] or super-resolution scanning microscopy [34-36].

## METHODS

**MCoSM setup and image acquisition**

Imaging was performed on a lab-built multi-color super-resolution microscope through modifying Nikon Eclipse Ti inverted microscope with a Nikon perfect focus system (PFS). The excitation sources were matched with the emission wavelengths $\lambda_1 = 405$ nm, $\lambda_2 = 488$ nm, and $\lambda_3 = 561$ nm, and $\lambda_4 = 647$ nm (OBIS 405, Sapphire 488, Sapphire 561, OBIS 647; Coherent). A spatial filter consisted of a lens with the focal length of 100 mm and a 5 μm pinhole (M-900, Newport). To maintain telecentricity, all distances between lenses are equal to the sum of their respective focal lengths [37]. The three synchronous output beams were combined using three beam splitter cubes (CCM1-PBS251/M, Thorlabs) and temporally synchronized using three spatial light modulators (SLM; PLUTO, Holoeye), where the output energy was ~50 mW.

Uploading different stacks phase distribution (corresponding with different wavelengths) on the SLM were adopted to realize super-resolution and phase-shifting in the focal plane, where the pixel size and pixel number are 8 μm×8 μm and 1080×1920, respectively. To ensure the same resolution (minimum step and the maximum range) in the $u$-plane and $v$-plane, the circular area with a diameter of 1080 pixels in the center was used. The beam modulated by SLM was expanded by the telescope to match the area of the entrance pupil of the illuminating objective. Two lenses were placed in a $4f$ configuration to image the phase distribution on the objective pupil to fill the entrance pupil of the microscope objective, where $f_{L1}=150$ mm and $f_{L2}=125$ mm, and the zeroth order of the beam was not included in the illumination path by mounted iris (ID8, Thorlabs). It should be noted that the amount of phase-shifting with multi-color illuminations were different, which was decided by the phase-shifting amount and incident wavelength.

In MCoSM, the same objective, excitation wavelength, and fluorescent filter were used both in the widefield and reconstructed images (Fig. S8). Removable mirrors (RM) 1 and 2 (BB05-E02, Thorlabs) were assisted by a removable bracket using an indexing mount (NX1N/M, Thorlabs), which was placed to determine the position of the sample using widefield illumination. The high-pass filter was coated by Cr with the thickness of 100 mm and the transmittance of OD3 (0.1%). Fluorescence was collected through an objective lens with the NA of 0.90 and magnification of $60^x$ (RMS60X-PFC, Olympus), a relay-system, and recorded on a 2048×2048 pixel sCMOS camera (Zyla 4.2 plus, Andor). The magnification of system $M_{sys}=50^x$ is determined by magnification of objective and $4f$ system. Realizing the adjustable spot array via mechanical-scan-free manner with the capture rates of up to 10 frames/s was used to ensure the delay reaction and fluorescence bleaching of sample. Although somewhat oversampled, we chose the above parameter to meet the Nyquist-Shannon sampling limit with 85 nm and reduce the acquisition time to ensure the performance of imaging. The scanning process with high-NA termed by the PFS, which was effectively ensured the fluorescent sample well in the experiment with 167.4 μm×167.4 μm FoV.

The prepared Lumisphere monodisperse microsphere solution is fluorescent beads (7-3-0010, Basel). Take 1 mL of fluorescent beads solution and drop it on a 0.17 mm cover glass and washed 4 times with alcohol buffer before imaging. The diameter of fluorescent beads is 100 nm with blue (350Ex/440Em), green (505Ex/515Em), red (580Ex/605Em), and cyan (633Ex/660Em) labeling. We demonstrated the capability of four-colors imaging through an objective lens with the NA of 1.25 and magnification of 60$^x$ (UPLFLN60XOI, Olympus) to extend the spatial and spectral performance.

We have also opted to study morphologically and ultrastructurally distinct cells of Bovine Pulmonary Artery Endothelial (BPAE) slide (F36924, Invitrogen Thermofisher) as a practical application in MCoSM. Nucleus (Cell-impermeant DAPI, 358Ex/461Em), cytoskeleton (Alexa Fluor 488 phalloidin, 505Ex/512Em), and mitochondria (MitoTracker Red CMXRos, 579Ex/599Em) of BPAE were imaging through an objective lens with the NA of 0.90 and magnification of 60$^x$ (RMS60X-PFC, Olympus).

**Phase distribution design**

Of particular interest for MCoSM is the design of the phase distribution of SLM that delivers super-resolution spot array illumination and phase-shifting scanning. We provide a modified iterative Fourier transform algorithm (IFTA) [38] to obtain super-resolution spot array and combining with diffractive phase shift grating to realize mechanical-scan-free imaging.

*A. Generation of adjustable diffractive spot array*

Increasing the number of illumination spots is desirable for improving FoV, however, achieving this capability in super-resolution spot array while maintaining good uniformity is challenging [39]. To obtain a super-resolution spot, the sampling interval in the focal plane has to reduce by applying a zero-padding operation to the input plane [25]. The size of the input plane zero-padding to at least 1/8 the size of the original one to precisely describe the details of the spot. The size of phase hologram is given approximately by $8M_x \times 8M_y$ according to the resampling, i.e. the size of the 50×50 (*N*=50) spot array can be two orders over the 3×3 (*N*=3) if the same precision sampling is used. Hence, a larger spot array exponentially increases the computational overhead leading to poor uniformity of spots during design, thereby limiting applicability in MCoSM settings. Here, we instead modified the algorithm by estimating the initial phase and adding an amplitude constraint strategy in the focal plane, and the working principle of the proposed iterative algorithm (Fig. 4a).

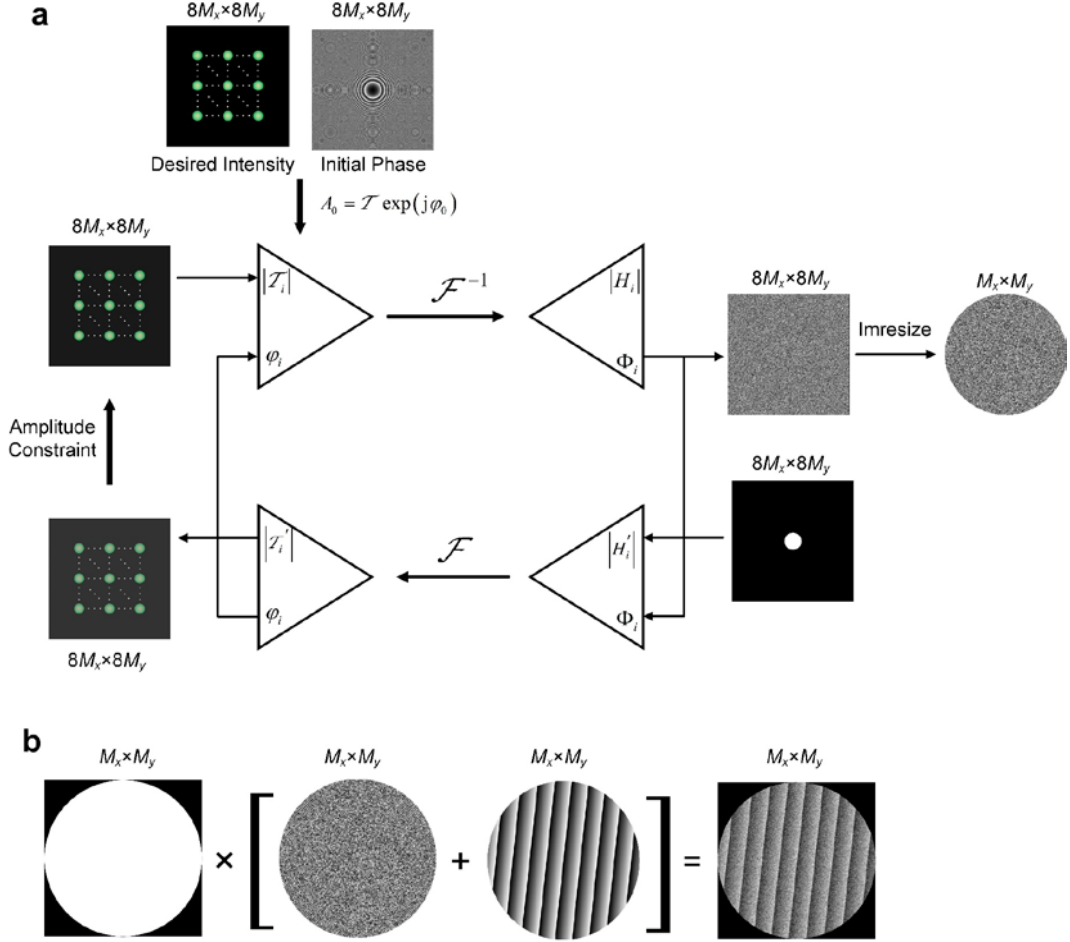

**Fig. 4. Process flow to produce a *N*×*N* super-resolution spot array with phase-shifting scan. a** Flowchart of the iterative algorithm for calculating phase distribution of hologram to generate a *N*×*N* super-resolution spot array. **b** Construction of hologram with spot array and phase-shifting scan.

Assuming that the resolution of the phase distribution is $M_x \times M_y$, the pixel size is $\Delta p$, and the size of the phase hologram is $a \times b$, where $a = M_x \Delta p$ and $b = M_y \Delta p$. During the phase hologram optimization, the pixel size of the hologram plane is not changed, while the size of the hologram plane is zero-padded to $8a \times 8b$ for reducing the sampling interval of the focal plane. The number of the sampling points in the hologram plane is changed as $8M_x \times 8M_y$, which is the same with the number of the sampling points in the focal plane.

The usual choice of the initial complex amplitude in the focal plane is $A_0(x,y) = \mathcal{T}_0(x,y)\exp\left[j\varphi_{\text{spot}}^{(0)}(x,y)\right]$ with a random phase $\varphi_{\text{spot}}^{(0)}(x,y)$. However, a random initial phase introduces many phase singularities, which would cause the stagnation problem of the iterative algorithm [40]. To avoid the stagnation while making the converges as quickly as possible, a two-dimensional estimated quadratic phase in the sample plane $(u,v)$ is used as the initial phase to approximately satisfy these conditions, which is expressed as

$$\varphi_{\text{spot}}^{(0)}(x,y) = C(x^2+y^2) , \qquad (5)$$

where $C$ is the positive coefficient, $x$ and $y$ are the coordinates of the focal plane.

Besides, the weighted constraint strategy is introduced into the iterative process to further enhance the result [41]. The focal plane is partitioned into two regions according to the intensity distribution of the desired intensity. The signal region is the area where the desired pattern is located, and the non-signal region is the area where there is no desired signal. During the iterative, the enforced amplitude constraint for the next iteration in the focal plane is

$$|A_i(x,y)| = \begin{cases} \mathcal{T}_i(x,y), & (x,y) \in S \\ |A'_{i-1}(x,y)|, & (x,y) \notin S \end{cases} , \qquad (6)$$

where $S$ denotes the signal region, and the amplitude field $\mathcal{T}_i(x,y)$ is replaced in each iteration, adopting negative feedback to improve the uniformity of the spot array, by the target amplitude by

$$\mathcal{T}_i(x,y) = \left[\sum_n \sqrt{\frac{\langle I_{i-1}(x,y)\rangle_N}{I_{i-1}(x_n,y_n)}}\right] \times \mathcal{T}_{i-1}(x,y) , \qquad (7)$$

where $\langle I_{i-1}(x,y)\rangle_N \approx (1/N)\sum_{n=1}^N I_{i-1}(x_n,y_n)$ and $\mathcal{T}_0(x,y)$ is the desired amplitude distribution, and the phase distribution remains unchanged. In $i$th iteration, the complex amplitude field $H_i = |H_i|\exp(j\Phi_i)$ in the input plane is calculated by the inverse Fourier transform of the complex amplitude field $A_i$ in the sample plane, and the amplitude distribution is replaced with the uniform amplitude after zero-padding [42]. This iterative process continues till the generated intensity profile in the focal plane converges to satisfy the constraints. The phase hologram can be generated from the complex amplitude field $H_i$ after a phase extraction and cropping operation. The amplitude constraint strategy in the focal plane improves the performance of the spot array with faster convergence by relaxing the amplitude constraint in the non-signal region (in Supplement note 4). It shows that the convergence of proposed method is fast and stable, and the RMSE and nonuniformity of our method significantly exceed the general IFTA, and the uniformity of the spot array is improved effectively.

The experimental result of the *N*×*N* spot array has presented in Fig. 5 with the NA of 0.9 at the incidence wavelength of 488 nm. The resolution of spot arrays was respectively 0.52Airy, 0.67Airy, 0.71Airy, 0.69Airy, and 0.68Airy, corresponding to Figs. 5a-5e. It can be seen from Fig. 5 that there are differences of intensity variation between ideal distribution and experimental results; however, the period of the spot array is not influenced.

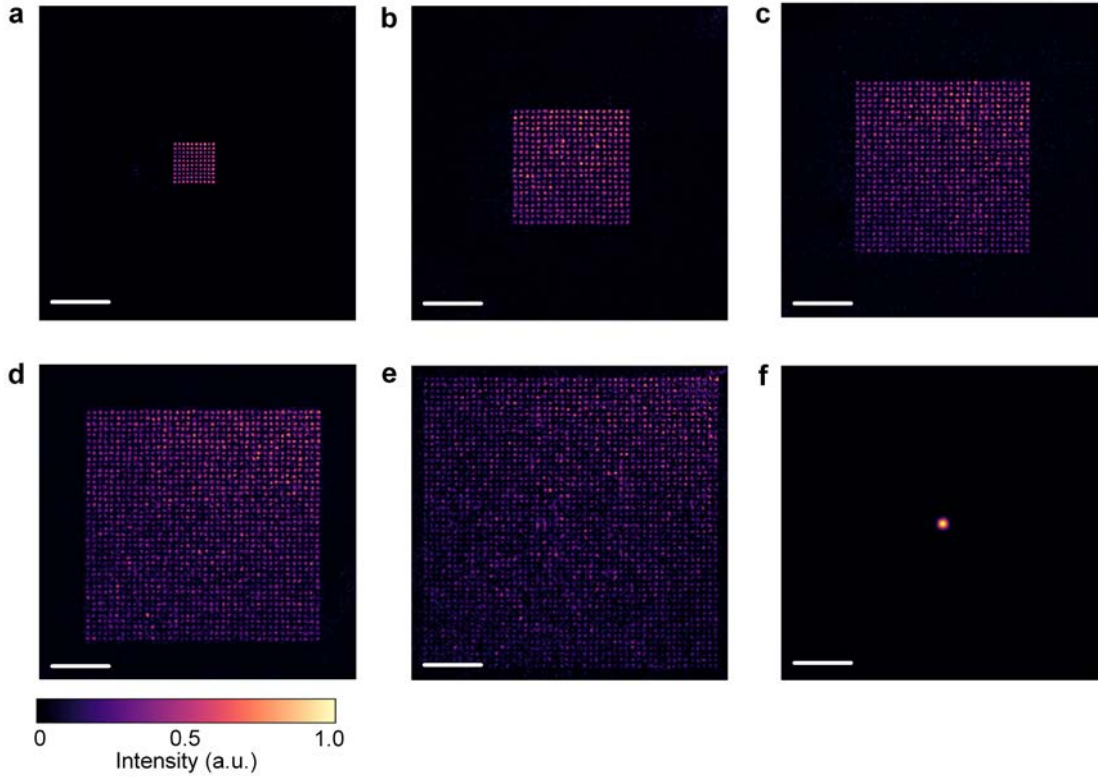

**Fig. 5. Experimental results of the *N*×*N* super-resolution spot array.** The results of **a** 10×10 (0.52Airy), **b** 20×20 (0.67Airy), **c** 30×30 (0.71Airy), **d** 40×40 (0.69Airy), **e** 50×50 (0.68Airy), and **f** Airy spot with the incidence wavelength of 488 nm at the NA of 0.90. Image size: 1500×1500 pixels. Scale bar: 1 μm.

*B. Generation of phase-shifting scanning*

To replace usual mechanical scanning manner, the phase-shifting is adopted to realize transverse scanning of spot array in the focal plane of the objective. According to diffraction equation, the relationship between the diffractive angle and wavelength can be expressed as [9]

$$2\Delta p \cdot \sin \Delta \theta = m\Delta\lambda, \qquad (8)$$

where $m=1$, $\theta$ is the obliquity of the added linear phase ramp, and $\Delta p$ is the pixel size of SLM. The distance between the center of the spot array and the focus of the Fourier lens is given by $\Delta l = f \cdot \tan\Delta\theta$, where $f$ is the focal length of the Fourier lens. Combining the distance with Eq. (8), the designed phase-shifting scanning can be described as

$$\varphi_{ps}(x, y) = [x\sin(\alpha) + y\cos(\alpha)]\tan\left[\arcsin\left(\frac{\Delta\lambda}{2\Delta p}\right)\right], \quad (9)$$

where $\alpha = \arcsin(x/\sqrt{x^2 + y^2})$.

C. Phase distribution combination

Combining the two phases distribution mentioned above, we add the phases of spot array and phase-shifting scanning to the uploaded phase distribution on the phase-only SLM. The phase loaded on the SLM is described as

$$\varphi_{SLM}(x, y) = \varphi_{spot}(x, y) + \varphi_{ps}(x, y). \quad (10)$$

The synthesis of the complete phase distribution displayed on the SLM (Fig. 4b). Meanwhile, the experimental results of 10×10 super-resolution spots with or without phase shift have been analyzed (Figs. S6b and S6c). It can be clearly seen that there is no significant difference in spot shape and intensity before and after loading phase shift (see Supplement note 3).

**Data processing**

A. Image pre-processing

Image pre-processing was divided into five steps. Obtained images were first batch cropped to remove phase-shifting scanning edge artifacts. Second, flat-field correction was then applied to obtained images prior to stitching to correct for illumination spot array inhomogeneity in the FoV. Third, applying a Gaussian kernel filter for each color to suppression stray light. Fourth, shake correction was necessarily in pre-processing to address dithering artifacts in liquid crystal of SLM and environmental vibrations that may cause phase shift inaccuracies. Fifth, for raster scanning imaging, the step size of phase shift was theoretically the same for each row of pixels, and dithering was corrected by averaging the step size of adjacent rows.

B. Super-resolution image reconstruction

To reconstruct the sample from the recorded intensity patterns, we correlate scanning data with registrations [43]. Eliminating drift over long periods acquisitions by pre-processing, we first calculated the average value of every 200 shots to generate a new stack. Second, the relative position of each spot can be determined to calculate the spot centers position and register each spot during the scanning process. By calculating the intensity of the spot centers, the scanned images of each spot can be obtained. Finally, the scanned images of all spots are merged into a whole image. Of note, the required width of each spot is $S=1.22\alpha \cdot \lambda \cdot M_{sys} / \Delta\mu \cdot \text{NA}$, where $\Delta\mu$ is the pixel size of detector.

*C. Multi-color stitching*

Excitation and detection efficiency usually depend on the position in the FoV and may differ by multi-color. To obtain multi-color image, the intensity should be corrected for colorimetric resolution analysis. Image analysis was performed using ImageJ (US, National Institutes of Health), Fiji (http://fiji.sc/), and MATLAB (MathWorks).

Multi-color stitching was performed using MATLAB and Fiji after color-balance correction [44]. Lateral position between color was evaluated using linear maximum-intensity projections to obtain optimum multi-color contrast [45, 46] where the interactions between different fluorescence labels was ignored. The reconstruction images were then divided by the normalized intensity profiles and merged into RGB composites.

**Funding.** This work was supported by the National Natural Science Foundation of China (Grants No. 62075112).

**Acknowledgments.** We thank Dr. Guoxuan Liu at Huawei Beijing R&D Centre for his helpful assistance during established the experiment and data processing. We also thank Dr. Calum Williams and Dr. Graham Spicer at the Cavendish Laboratory, University of Cambridge, UK for their helpful suggestions during manuscript preparation.

**Disclosures.** The authors declare no competing interests.

**Data availability.** The data that support the findings of this study are available from the corresponding author upon request.

**Code availability.** Custom codes used for analysis and image processing pipelines are available from the corresponding author upon request.

**Supplemental document.** See Supplement note for supporting content.